\documentclass[conference]{IEEEtran}
\IEEEoverridecommandlockouts

\usepackage{graphicx}
\usepackage{textcomp}
\usepackage{xcolor}
\usepackage{textcase}
\usepackage[]{caption}
\usepackage[normalem]{ulem}
\usepackage{subcaption}
\useunder{\uline}{\ul}{}
\usepackage{multirow}
\usepackage{multicol}
\usepackage{amsmath}
\usepackage{booktabs}
\usepackage[paper=letterpaper, top=0.75in, bottom=0.75in, left=0.75in, right=0.75in]{geometry}
\usepackage{pifont}
\usepackage{hyperref}
\newcommand{\cmark}{\ding{51}}%
\newcommand{\xmark}{\ding{55}}%

\usepackage{cite}
\usepackage{amsmath,amssymb,amsfonts}
\usepackage{algorithmic}
\usepackage{graphicx}
\usepackage{textcomp}
\usepackage{xcolor}
\def\BibTeX{{\rm B\kern-.05em{\sc i\kern-.025em b}\kern-.08em
    T\kern-.1667em\lower.7ex\hbox{E}\kern-.125emX}}
    
\begin{document}

\newgeometry{paper=letterpaper, top=0.75in, bottom=0.75in, left=0.75in, right=0.75in, includehead}

\title{\LARGE \bf Improving Continuous Grasp Force Decoding from EEG with Time-Frequency Regressors and Premotor-Parietal Network Integration}

\author{ Parth G. Dangi$^{1}$, Yogesh Kumar Meena$^{*1}$
\thanks{$^{1}$ Parth G. Dangi and Yogesh Kumar Meena are with Human-AI Interaction (HAIx) Lab, IIT Gandhinagar, India
        {\tt\small yk.meena@iitgn.ac.in}}%
}


\maketitle

\begin{abstract}

Brain-machine interfaces (BMIs) have significantly advanced neuro-rehabilitation by enhancing motor control. However, accurately decoding continuous grasp force remains a challenge, limiting the effectiveness of BMI applications for fine motor tasks. Current models tend to prioritise algorithmic complexity rather than incorporating neurophysiological insights into force control, which is essential for developing effective neural engineering solutions. To address this, we propose \textit{EEGForceMap}, an EEG-based methodology that isolates signals from the premotor-parietal region and extracts task-specific components. We construct three distinct time-frequency feature sets, which are validated by comparing them with prior studies, and use them for force prediction with linear, non-linear, and deep learning-based regressors. The performance of these regressors was evaluated on the WAY-EEG-GAL dataset that includes 12 subjects. Our results show that integrating \textit{EEGForceMap} approach with regressor models yields a 61.7\% improvement in subject-specific conditions (R² = 0.815) and a 55.7\% improvement in subject-independent conditions (R² = 0.785) over the state-of-the-art kinematic decoder models. Furthermore, an ablation study confirms that each preprocessing step significantly enhances decoding accuracy. This work contributes to the advancement of responsive BMIs for stroke rehabilitation and assistive robotics by improving EEG-based decoding of dynamic grasp force.

\end{abstract}


\section{Introduction}

\IEEEPARstart{T}he number of stroke patients and amputees is rising significantly worldwide, particularly among younger populations. This trend is more prominent in low- to middle-income countries, where upper limb amputations and stroke-related conditions are steadily increasing~\cite{{Kurucan2020-pc},{dar2024outcome}}. 
Populations affected by stroke and amputations from these countries require critical attention during their rehabilitation and restoration of upper limb movement, which is available to the same demographic in high-income countries~\cite{Kurucan2020-pc}. 
Multiple approaches such as chemical interventions and physical exercises have been proposed and implemented to achieve this goal~\cite{Mehrabi2024-zr}. Out of which, brain-machine interface (BMI) technology is one of the most promising ways by which upper-limb movement can be restored~\cite{hortal2015using}.

BMI involves systems (hardware and software) that allow humans to interact with their surroundings without involving peripheral nerves and muscles by using control signals generated from brain activity~\cite{ng}. BMI has been shown to restore the upper-limb movement in stroke patients as well as amputees during rehabilitation~\cite{{ng}}. Within the BMI models, the motor imagery based brain-computer interfaces (MI-BCI) have shown significant success in restoring the upper-limb movement of their users~\cite{ng, {chowdhury2018active},{chowdhury2019eeg}}. The MI-BCI apparatus records the brain activity occurring during the MI process -- the visualisation and planning of the movement occurring before/in the absence of the movement execution~\cite{j94}, and converts it into mechanical or computer hardware commands.

The earliest successful use of the MI-BCI system can be traced back to Wolpaw et al.~\cite{wm04}, who developed a BCI system performing reaching tasks in a virtual environment through imagined reach and grasp movement. This event catalysed the rapid development of MI-BCI systems, especially for accurately decoding continuous motion features, such as trajectory~\cite{ng}, speed~\cite{osvimbrfa24} and, direction~\cite{10831499}, by using single modality~\cite{10.1007/978-981-99-8021-5_7} and hybrid BCIs~\cite{6999180, 7318410, pfurtscheller2010hybrid, meena2015simultaneous}. However, decoding kinetic features such as force was limited, as earlier studies suggested force is encoded discretely~\cite{m89}. Consequently, earlier kinetic decoders classified force into subjective categories. Recent findings, however, show force is encoded within a defined premotor-parietal network~\cite{ds09}. 

The discovery of the encoding mechanisms for force control~\cite{ds09} has opened up the possibility of decoding features that were previously considered discrete, such as force~\cite{m89}, in a continuous and accurate manner. To investigate this potential, several approaches have been proposed in the last decade for decoding grasp force directly from brain activity. Initial studies, such as the one by Flint et al.~\cite{fo14}, utilised semi-invasive imaging techniques like electrocorticography (ECoG). However, this method resulted in limited applications for the force decoders. To expand the application of force decoding algorithms, researchers have developed decoders that use non-invasive imaging methods like electroencephalography (EEG)~\cite{10.1007/978-981-99-8021-5_7}. However, inherently weak EEG signals, due to low signal-to-noise ratio and spatial ambiguity~\cite{fatourechi2007emg}, have led previous studies to combine them with electromyography (EMG) muscle activity signals to improve accuracy~\cite{luciw2014multi}. Unfortunately, this approach limits decoders to executed movements, restricting their use in neuroprosthetics and early stroke rehabilitation, where decoding from MI signals is essential~\cite{{ng}}.
\restoregeometry

\begin{figure*}[ht!] 
\centering
\includegraphics[width=1.0\linewidth]{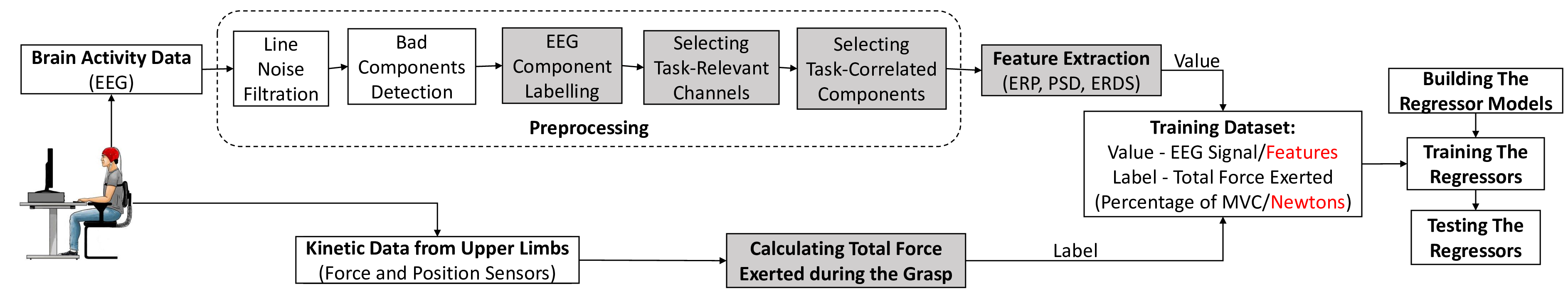}
\caption{Schematic representation of the pipelines employed for decoding exerted force. The grey boxes and the red text in the white boxes show new modifications that have been introduced by \textit{EEGForceMap}.}
\label{Figure 1}
\end{figure*} 

Decoding grasp force from single-modality EEG data has gained attention, leading to the proposal of several efficient approaches in regressor model, such as using deep-learning models inspired by spike activity in the brain~\cite{kumarasinghe2021brain} and developing ensembles of linear classifiers and deep-learning based decoders to classify the brain activity data into small subsets, followed by decoding grasp force from each of these subsets~\cite{10.1007/978-981-99-8021-5_7}. However, these methods often lack a well-developed preprocessing and feature extraction pipeline, which is essential for effectively extracting relevant information from EEG data. As a result, the complex regressor models developed may not provide clear insights into the nature of the EEG signals. This can lead to overfitting~\cite{kumarasinghe2021brain}, particularly for subject-specific data, as well as a limited understanding of the EEG data itself~\cite{10.1007/978-981-99-8021-5_7}. In this work, we address these challenges by proposing an approach that reintroduces insights from neural force control into the grasp force decoding process. The main contributions of this work are:

\begin{itemize}

    \item Introducing the \textit{EEGForceMap} approach that enhances EEG preprocessing by selecting task-specific channels and components from the premotor-parietal network.

    \item Presenting a wide range of time-frequency domain feature sets (ERP, PSD, and ERDS) for decoding exerted grasp force, validated by prior fMRI studies on neural mechanisms underlying force control. 

    \item Developing and evaluating classical regression-based models for decoding exerted force from EEG features, comparing their performance with state-of-the-art methods using a publicly available WAY-EEG-GAL dataset.  

\end{itemize}

\section{Proposed Methodology} \label{Section 3}

We developed the \textit{EEGForceMap} methodology to decode finger force exertion from brain activity recorded via EEG. As shown in Figure \ref{Figure 1}, the EEGForceMap pipeline consists of key steps, including data preprocessing, feature selection/extraction, and decoding models.

\subsection{Data Preprocessing}


The first part of the proposed methodology involves an EEG preprocessing pipeline designed to filter brain activity signals and select relevant EEG signals from the parietal and premotor cortex, which are associated with kinetic force control~\cite{Ehrsson2001-nf, ds09}. Prior to data analysis, the EEG signals were subjected to a notch filter with a frequency range of 0.5 to 50 Hz. To further refine the data, EEG components were separated, and those exhibiting abnormal patterns were removed. These techniques have been employed in previous studies to eliminate instrumental errors in EEG data~\cite{10.1007/978-981-99-8021-5_7, luciw2014multi}. To enhance the EEG data quality, the following new steps were introduced in the \textit{EEGForceMap} preprocessing pipeline. 


\subsubsection{EEG Component Labelling} Independent Component Analysis (ICA) was applied to decompose the EEG signal. Components that exhibited line noise were identified and removed. The ICLabel library used deep-learning algorithms to label muscle, eye movement, and cardiac artifacts by comparing these components with patterns seen in previous studies~\cite{PIONTONACHINI2019181}. These non-neural components were excluded to retain brain activity. This method has been applied in both offline and pseudo-online studies~\cite{9967664}, but its effectiveness in online trials has not been tested previously, making this one of the novel aspects of this work.


\subsubsection{Selecting Task-Relevant Channels} After component labelling, a specific number of channels were selected to isolate the task-relevant EEG data while reducing the load on the preprocessing pipeline. The EEG channels were selected based on their covariance with grasp force to optimise feature extraction and decoding, while still retaining essential information. To identify channels that needed to be selected, a small subset of EEG data from each subject was manually analysed to measure covariance with the grasp force,1 which has been shown to be an effective selection strategy than other methods~\cite{PY2019}. These covariance values were then compared to identify which EEG channels were most sensitive to changes in grasp force. Ultimately, the channels 'C3', 'CP1', 'CP2', 'Cz', 'FC2', 'FC6', 'Fp1', 'P3', and 'C4' demonstrated the highest sensitivity to changes in grasp force and hence,  were selected to provide concise changes in the EEG data during preprocessing.

\subsubsection{Selecting Task-Correlated Components} Finally, the EEG components from the selected data were further refined. To achieve this, all EEG components from a small subset of EEG data from each subject were analysed for their covariance with grasp force offline. Each component's covariance was then computed and arranged in ascending order. The components with the highest covariance were identified and isolated to reconstruct an EEG signal that is more sensitive to grasp force changes. After these steps, we obtained the denoised and selected EEG data, which contained information specific to grasp force. This data was used for decoding and feature extraction.

\subsection{Feature Extraction}

After preprocessing, time-frequency domain features were extracted from the selected EEG data using a sliding window approach (100 ms window size, 50 ms step) to enable real-time processing. This method was chosen to reduce noise and minimise errors from individual EEG spikes. Feature selection was guided by similarities with fMRI studies~\cite{Ehrsson2001-nf}. Three feature sets were developed: event-related potentials (ERP), power spectral density (PSD), and event-related desynchronisation/synchronisation (ERDS), aligning with state-of-the-art methods~\cite{{osvimbrfa24}, {kssc14}, {nsthk14-1}}.

\subsubsection{Event-related Potentials} The ERP feature set was designed to efficiently extract features from small windows of EEG data. To reduce the computation cost during feature extraction and decoding, the ERPs were simplified using statistical parameters, including mean, mean absolute value, area under the curve (AUC), skewness, kurtosis, and variance, to describe these ERPs accurately. While previous studies applied these parameters for force classification~\cite{osvimbrfa24}, we used them for force decoding. However, this feature set was found to be highly dimensional. To address this high dimensionality, principal component analysis (PCA) was employed, enabling us to reduce the number of dimensions while retaining 95\% of the variance.


\subsubsection{Power Spectral Density} PSD feature set explicitly captures the magnitude of activations in each frequency band as PSD. The Mu (9-11 Hz) and Theta (4-8 Hz) bands, which are associated with the premotor-parietal network of movement planning \cite{ds09}, were of particular interest. Thus, the PSD activity from these bands was calculated as follows: 
\begin{equation}
\text{$PSD$}(f) = \lim_{T \to \infty} \frac{1}{T} \left| X_T (f) \right|^2
\end{equation}
where $PSD (f)$ is the PSD of the signal $X(t)$ at frequency band $f$. $T$ is the window size, and $X_T (f)$ is the Fourier transform of the signal in the interval $[-T/2, T/2]$. 

\subsubsection{Event-related Desynchronisation/Synchronisation} The ERDS feature set was designed to capture the relative change in the power of the EEG signal during a specific task at a given time by calculating ERDS of the EEG data. While the ERDS has been used in prior decoder models~\cite{nsthk14-1} to classify force into discrete levels with high accuracy, its application in force decoding has not yet been tested. Therefore, we employed this feature set to decode the exerted force. A key challenge in calculating ERDS was calculating the baseline, which was established by correlating PSD values with grasp force levels, specifically when total load force was below 0.5 N. ERDS was calculated for Mu and Theta bands as:
\begin{equation}
\text{$ERDS$} = \frac{\text{$PSD$}_{\text{$active$}} - \text{$PSD$}_{\text{$baseline$}}}{\text{$PSD$}_{\text{$baseline$}}} \times 100
\end{equation}
where $PSD_ {active}$ represents the PSD values recorded at that time, while $PSD_ {baseline}$ refers to the baseline PSD. The ERDS values were explicitly calculated for the Mu and Theta bands, which were previously used to construct the PSD features.

\subsection{EEG Regressor Models}

The final step of the \textit{EEGForceMap} methodology is to integrate the regressor models to decode exerted force from extracted features. For this, we integrated linear~\cite{fo14}, non-linear~\cite{pgpsc19}, and deep learning~\cite{10.1007/978-981-99-8021-5_7} regressors with the \textit{EEGForceMap}. These models were selected because of their prior use in multiple state-of-the-art (SOTA) decoding approaches and their lower computation cost and simpler architecture. These models are:

\subsubsection{Linear Regression Models} 

We implemented two classical linear regression models to decode exerted force from brain activity: the single feature linear regressor (SFLR) and the multiple linear regressor (MLR). The SFLR utilised only the statistical feature set for decoding, whereas the MLR leveraged all three feature sets. Both models performed linear regression as follows:

\begin{equation}
Y = \beta_0 + \sum_{i=1}^n \beta_i X_i + \epsilon
\end{equation}

Where \(\beta\) represents the regression model weights, \(X\) denotes the feature matrices, and \(\epsilon\) accounts for the error term. 

\subsubsection{Partial Least Squares Regression Models} Partial least squares regressor (PLSR) was implemented to decode force using all feature sets. PCA was applied to reduce dimensionality while preserving variance. The model was optimised through hyperparameter tuning using a grid search with cross-validation over 20 components:
\begin{equation}
\hat{\mathbf{y}} = \mathbf{X}_{\text{PLS}} \mathbf{B}
\end{equation}
Where $\hat{y}$ denotes the predicted grasp force, $\mathbf{X}_{\text{PLS}}$ denotes the partial least squares of the input data, and $\mathbf{B}$ is the error. This process optimised the number of components, minimised mean squared error (MSE), and enhanced model consistency.

\subsubsection{Neural Network Regression Model} A neural network regressor (NNR) was implemented with an architecture consisting of an input layer, three hidden layers (8, 16, 32 neurons), and an output layer. It mapped ERDS features to continuous force values. Each hidden layer transformed inputs as:
\begin{equation}
\mathbf{z}^{(l)} = (\mathbf{W}^{(l)} \mathbf{z}^{(l-1)} + \mathbf{b}^{(l)}), \quad l = 1, 2, 3,
\end{equation}
Where ${W}^{(l)} \in \mathbb{R}^{m_l \times m_{l-1}}$ is the weight matrix, and ${b}^{(l)} \in \mathbb{R}^{m_l}$ is the bias vector of the \(l\)-th layer. The output layer follows a similar architecture, providing the predicted force value. The model was optimised to minimise MSE loss using the Adam optimiser:
\begin{equation}
\mathcal{L} = \frac{1}{n} \sum_{i=1}^n \left( y_i - \hat{y}_i \right)^2
\end{equation}
Where \(y_i\) and \(\hat{y}_i\) are the actual and predicted values of the load force for the \(i\)-th sample, respectively. The optimisation process uses the Adam optimiser, which updates the weights iteratively to minimise \(\mathcal{L}\) with adaptive learning rates. 

\section{Experimental Protocol}

The performance of \textit{EEGForceMap} was evaluated using a publicly available dataset\cite{luciw2014multi} to decode force exerted during grasp-and-lift tasks from EEG brain activity.

\subsection{EEG Data Preparation}



For performance evaluation, the grasp-and-lift trial data was taken from the WAY-EEG-GAL dataset prepared by Luciw et al.~\cite{luciw2014multi}. This dataset comprises multi-channel EEG recordings from 12 healthy, right-handed individuals performing grasp-and-lift trials designed to study the neural mechanisms underlying movement planning~\cite{luciw2014multi}. The experimental task involved lifting objects with varying weights and surface friction. From these trials, EEG data were recorded using a 64-channel system at a sampling rate of 2,048 Hz, accompanied by EMG, kinematic, and kinetic data to capture muscle activity, motion, and forces during the task~\cite{luciw2014multi}. 

To accurately assess grasp force prediction, only trials with constant surface and varying force levels were used. EEG and kinetic data were extracted from these trials, retaining only channels with force sensor information. For subject-specific models, each participant's data were randomly shuffled and split into training, testing, and validation sets (7:2:1). For the subject-independent model, the last participant's data were reserved for testing, whereas the remaining data were used for training.

\subsection{Evaluation Setup}


To evaluate and assess \textit{EEGForceMap} methodology more effectively, all the regressor models were trained under subject-specific and subject-independent conditions. The subject-specific models were tailored and tested for each individual participant. In the subject-independent setup, one participant was designated for testing while the remaining participants were utilised for training. This approach resulted in 12 different models, each with a unique participant as the test subject.

Before decoding, the total grasp force was computed by summing the force sensor values from both fingers at each timestamp. To simulate real-world data collection, the \textit{EEGForceMap} methodology was trained under pseudo-online conditions using a sliding window of 0.1 seconds with a 0.05-second step size, which has been used in previous studies~\cite{10.1007/978-981-99-8021-5_7}. The training and validation sets were derived from the grasp-and-lift trial dataset. Each regressor was trained and evaluated separately for each participant to account for inter-subject variability. Each model was trained for 100 epochs. The experiment was conducted on a Dell Inspiron 15 3000 computer with an Intel(R) i3-1115G4 processor and 16GB RAM.

\subsection{Performance Matrices}

Finally, each of the regressors integrated with \textit{EEGForceMap} was evaluated based on its performance. The performance of each regressor was analysed by (1) the correlation of feature sets with grasp force and (2) decoding performance using four regressors. Feature accuracy was measured via Pearson's correlation coefficient (Pearson's C). When Pearson's C values were similar across sets, the coefficient of variation ($CV$)—the ratio of standard deviation to mean—was used to assess sensitivity to grasp force changes. Decoder performance was evaluated using the coefficient of determination ($CoD$), a standard metric for SOTA models~\cite{pgpsc19, fu2023eeg}, computed per participant as:

\begin{equation}
CoD = 1 - \frac{\sum_{i=1}^N (Y_i - \hat{Y}_i)^2}{\sum_{i=1}^N (Y_i - \bar{Y})^2}
\end{equation}

where \( Y_i \) is the actual load force, \( \hat{Y}_i \) the predicted force, \( \bar{Y} \) the mean actual force, and \( N \) the number of samples.


Subject-specific model $CoD$ values were compared across different regressors to assess decoding performance. These values were also benchmarked against SOTA methods using the same WAY-EEG-GAL dataset under subject-specific conditions. Following this, the performance of subject-independent models was compared to that of subject-specific models using a two-way ANOVA. This comparison aimed to confirm that the decoders effectively isolate brain activity associated with force control for grasp force decoding.

\subsection{Ablation Study}

To evaluate the effectiveness of the proposed \textit{EEGForceMap} methodology, we conducted an ablation study using ANOVA with a 90\% confidence level. This involved systematically removing each of the four preprocessing steps: line noise filtering, task-related component selection, task-related channel selection, and feature extraction (see Fig.~\ref{Figure 1}). For feature extraction, we employed a feature set specifically designed to capture grasp force characteristics. The final decoding was performed using the best-performing regressor model identified in our experiments. The ablation study was carried out using one run per participant from the WAY-EEG-GAL dataset~\cite{luciw2014multi}, comprising 10 minutes of EEG data and 12 grasp-and-lift trials across three weight conditions (165gm, 330gm, and 660gm), to account for inter-subject variability.


\section{Results and Analysis}

\subsection{The Correlation of Extracted Features}

\begin{figure*}
    \centering
    {\includegraphics[width=0.37\textwidth]{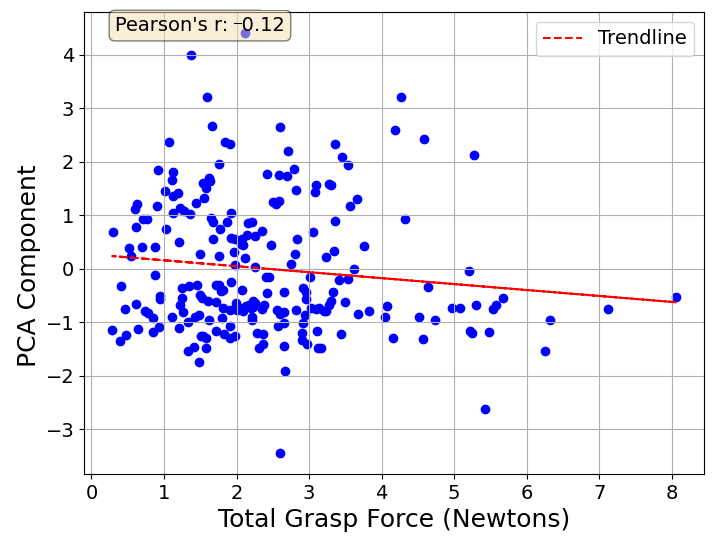}} 
    \hspace{5mm}
    {\includegraphics[width=0.38\textwidth]{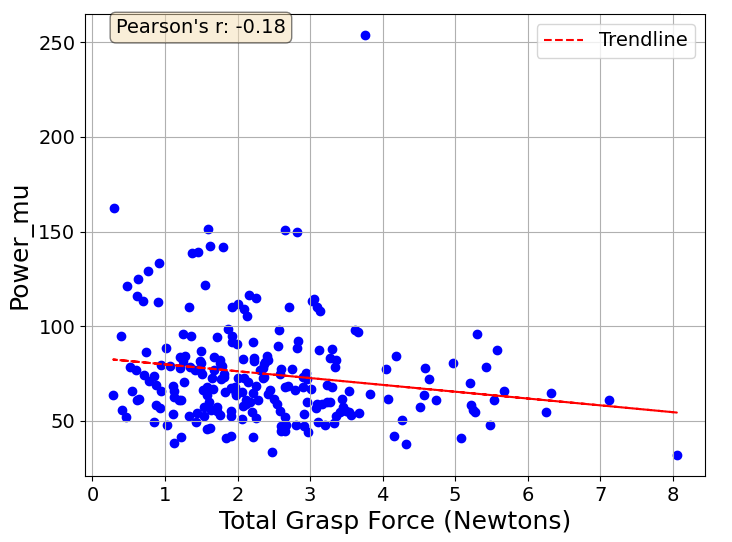}} 
    \caption{ Correlation between ERP (Left) and PSD features (Right) extracted using \textit{EEGForceMap}, and exerted grasp force.}
    \label{Figure 2}
\end{figure*}

\begin{figure}
\centering
\includegraphics[width=0.4\textwidth]{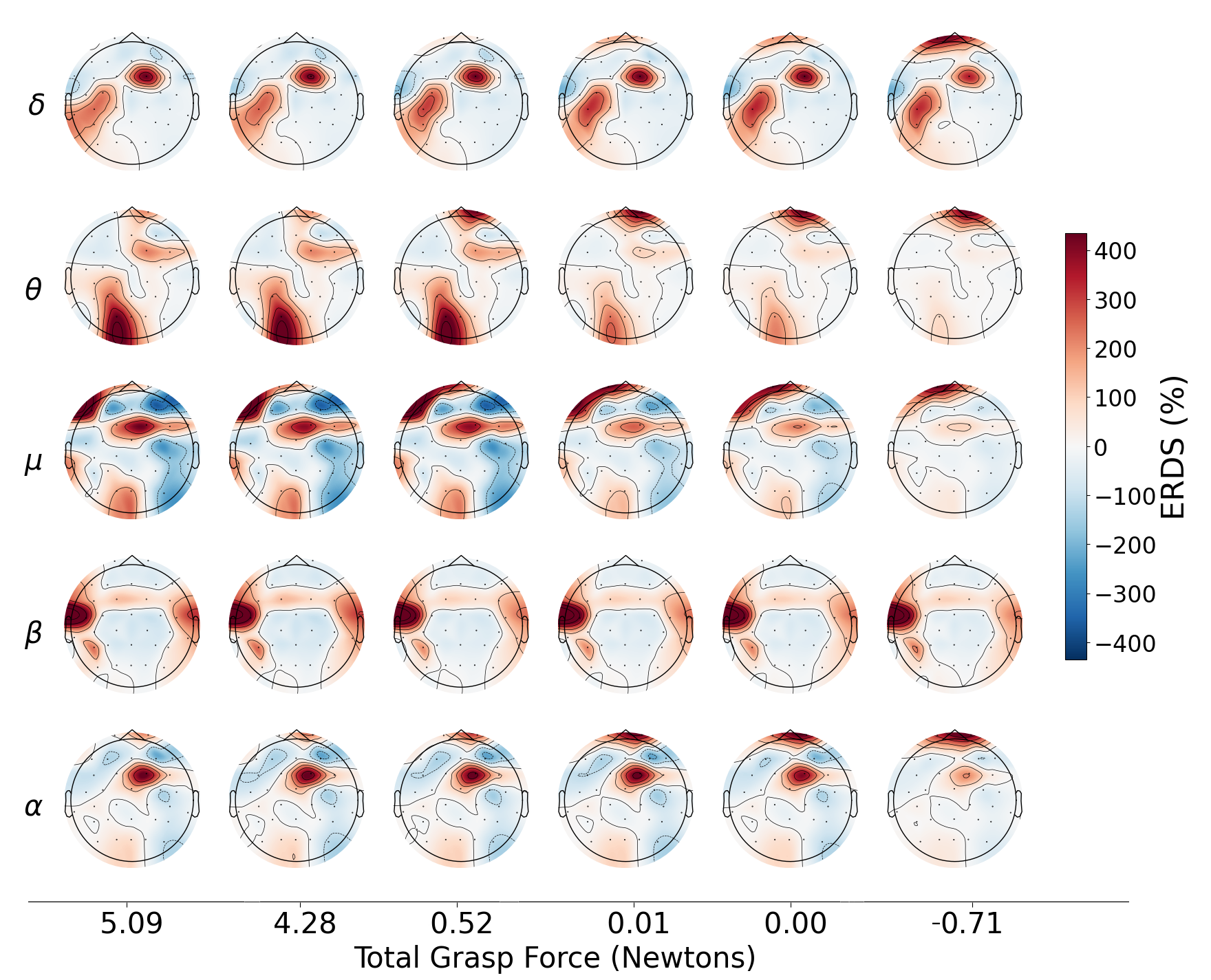}
\caption{Spatial representation of the changes seen in the ERDS features at different bands and grasp force.}
\label{Figure 3}
\end{figure}


To assess \textit{EEGForceMap}'s performance across regressors, we examined how each feature set contributed to decoding. Correlating each set with grasp force revealed that ERP features had the lowest correlation (Pearson's C = 0.12), contrary to prior findings~\cite{osvimbrfa24}. As shown in Fig.~\ref{Figure 2} (left), multiple ERP feature values are mapped to similar force levels. This suggests their generalised representation of brain activity—suitable for classification—may lack the detail needed for accurate force decoding.

Meanwhile, PSD features moderately correlated with grasp force (Mu band: Pearson's C = 0.28, Theta band: Pearson's C = 0.34). Fig.~\ref{Figure 2} (right) shows feature values corresponding to specific force levels, likely due to the emphasis on brain activity magnitude during extraction. However, high redundancy was observed, with multiple feature values mapping to similar force ranges. Cross-referencing with force data revealed these redundancies as EEG deviations captured during extraction, reducing sensitivity to force variations ($CV$ = 32.41).

This deviation was significantly reduced in ERDS features by normalising PSD changes against a baseline, enhancing sensitivity to force ($CV$ = 356.65; t-score = -10.3659, p $\le$ 0.000001), despite similar correlations (Mu: C = 0.26, Theta: C = 0.32). ERDS features amplified activation in parietal and premotor regions (Fig.~\ref{Figure 3}), aligning with prior fMRI findings~\cite{{ds09},{Ehrsson2001-nf}}. This, in total, shows that the ERDS features encode the grasp force, and hence can be used as input for regressors as well as for the ablation study.

\begin{figure}
\centering
\includegraphics[width=0.43\textwidth]{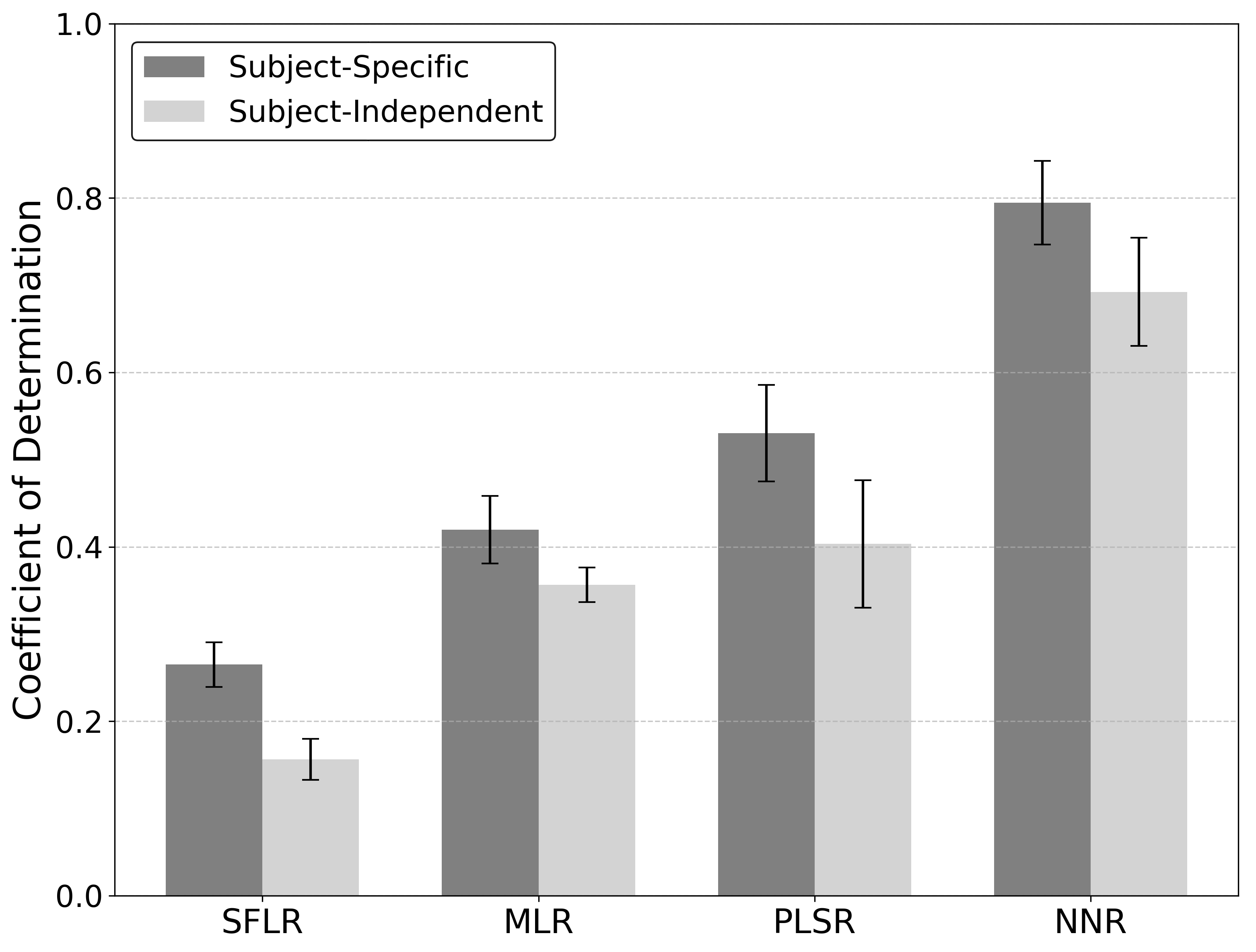}
\caption{Average decoding accuracies of regressor models integrated with the proposed \textit{EEGForceMap} methodology.}
\label{Figure 4}
\end{figure}


\subsection{Comparison of Decoder Models}

\begin{table*}[]
\centering
\caption{A comparison of the coefficient of determination (CoD) between subject-specific (SS) and subject-independent (SI) models, using the \textit{EEGForceMap} pipeline, with SOTA methods for decoding grasp force from EEG signals in human subjects.}
\label{Table 2}
\begin{tabular}{llllll}
\hline
\multirow{2}{*}{Kinetic Decoding Models} &
  \multirow{2}{*}{Model Type} &
  \multirow{2}{*}{Data  Type} &
  \multirow{2}{*}{Decoding Approach} &
  \multicolumn{2}{c}{CoD} \\ \cline{5-6} 
                                                                  &          &           &               & SS & SI \\ \hline
Multi-Modal Decoder~\cite{fu2023eeg}                              & CNN-LSTM & EEG + EMG & Deep-Learning & 0.457       & -           \\
Classifier-Decoder  Ensemble Model~\cite{10.1007/978-981-99-8021-5_7} &
  LDA + CNN-LSTM &
  EEG &
  Ensemble Approach &
  0.487 &
  - \\
Brain-Inspired  Spike Neural Network~\cite{kumarasinghe2021brain} & SNN      & EEG       & Deep-Learning & 0.504       & -           \\
\textit{EEGForceMap} + SFLR                                       & LR       & EEG       & Linear        & 0.261       & 0.156       \\
\textit{EEGForceMap} + MLR                                        & MLR      & EEG       & Linear        & 0.425       & 0.356       \\
\textit{EEGForceMap} + PLSR                                       & PLSR     & EEG       & Non-Linear    & 0.536       & 0.405       \\
\textit{EEGForceMap} + NNR                                        & MLP      & EEG       & Deep-Learning & 0.815       & 0.785       \\ \hline
\end{tabular}
\end{table*}

\begin{table*}[]
\centering
\caption{Ablation study showing the effect of each preprocessing step on the coefficient of determination (CoD).}
\label{Table 3}
\begin{tabular}{cccc|l|l}
\hline
Line  Noise Filter &
  \begin{tabular}[c]{@{}c@{}}Selecting Task-\\ Related Components\end{tabular} &
  \begin{tabular}[c]{@{}c@{}}Selecting Task-\\ Related Channels\end{tabular} &
  Feature  Extraction &
  \multicolumn{1}{c|}{\begin{tabular}[c]{@{}c@{}}Input EEG \\ Data Status\end{tabular}} &
  \multicolumn{1}{c}{CoD} \\ \hline
\xmark & \xmark & \xmark & \xmark & Raw Data               & 0.236 \\
\cmark & \xmark & \xmark & \xmark & Filtered Raw Data      & 0.265 \\
\cmark & \cmark & \xmark & \xmark & Reconstructed EEG Data & 0.333 \\
\cmark & \cmark & \cmark & \xmark & Selected EEG Data      & 0.407 \\
\cmark & \cmark & \cmark & \cmark & Features from EEG Data & 0.785 \\ \hline
\end{tabular}
\end{table*}

After evaluating the contributions of each feature set to the decoding process, we assessed the decoding performance of each model by comparing the $CoD$ across all subjects using two-way ANOVA. The results indicated significant differences in the $CoD$ among the proposed models (F-value: 7.4049; p-value = 0.0006), with no significant inter-subject variation ($f$-value: 0.7991, $p$-value = 0.64). However, we saw that the subject-specific models were significantly more accurate than subject-independent models ($f$-score = 3.512, $p$-value = 0.012; See Fig.~\ref{Figure 4}), However the effect size for this observation is less (Cohen's D: 0.182), suggesting that this effect is minor and caused due to the individual differences caused by subject-specific patterns. When comparing \textit{EEGForceMap}'s regressors with SOTA methods, we found that our models achieved higher decoding accuracy. The performance of each regressor model is discussed below. 


\begin{figure}[h!] 
\centering
\includegraphics[width=0.44\textwidth]{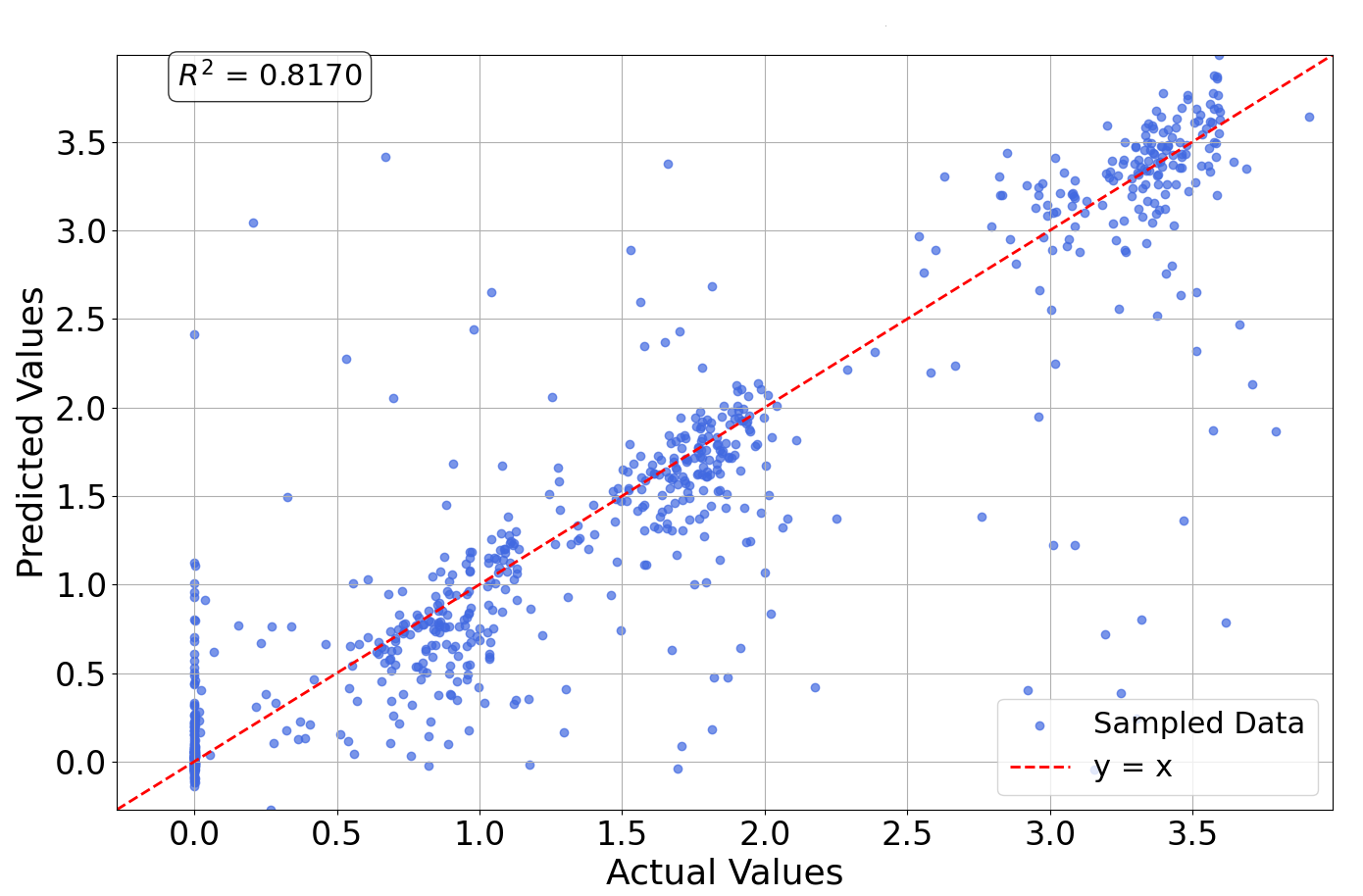}
\caption{Comparison of actual and predicted grasp force to assess NNR model
in predicting exerted grasp force.}
\label{Figure 5}
\end{figure}

\subsubsection{Linear Regressors}

SFLR and MLR are basic EEG decoding method, commonly used in earlier work~\cite{pgpsc19}. SFLR uses a single feature set and shows the lowest accuracy in both our pipeline and compared SOTA models (Table~\ref{Table 2}) in both subject-specific and subject-independent conditions (See Fig.~\ref{Figure 4}). MLR, by integrating all feature sets, significantly improves performance ($t$-score: 2.94, $p$-value = 0.02). Even though the MLR's performance is better than its predecessor, its performance accuracy is still lower than that of the SOTA models in subject-specific and subject-independent models (see Fig.~\ref{Figure 4}). 


\subsubsection{Partial Least Square Regressor}

PLSR, a non-linear method often used for kinematic decoding~\cite{kssc14}, outperforms linear models in accuracy ($t$-score: 1.943, $p$-value = 0.005), as shown in Fig.~\ref{Figure 4}. When compared to the SOTA, the PLSR shows significantly improved performance accuracy. However, it exhibits significantly higher inter-subject variation (IQR = 0.36) than the SOTA and the proposed models. This was explicitly seen when the subject-specific and subject-independent PLSR models showed no significant difference ($t$-score = 1.2364, $p$-value = 0.193, see Fig.~\ref{Figure 4}), reducing its generalisability required for real-time decoding.

\subsubsection{Neural Network Regressor}

The NNR model is a deep-learning approach widely used in BCI models~\cite{fu2023eeg, pgpsc19}. This model shows outstanding performance at decoding grasp force (see Table~\ref{Table 2}). These models have shown greater decoding accuracy than both SOTA and proposed models in subject-specific and subject-independent settings (see Fig.~\ref{Figure 4}). However, this model is not able to distinguish between randomly produced force during the rest period and initial force buildup during the grasp-and-lift trials, leading to significant error patterns as described in Fig.~\ref{Figure 5}.

\subsection{Ablation Study Outcomes}

The ablation study was conducted to critically analyze the impact of each step of the \textit{EEGForceMap} on its performance. Our findings indicate that removing the ERDS feature extraction (FE) significantly decreased accuracy ($f$-score = 3458.53; $p$-value $<$ 0.00001), highlighting its importance. In contrast, component selection slightly improved performance ($f$-score = 3.149; $p$-value = 0.08), suggesting potential for further improvements through dynamic selection.

We also discovered that channel selection only enhanced accuracy when feature extraction was included. (Without FE: $f$-score  = 1.319, $p$-value = 0.2699; With FE: $f$-score  = 128.54, $p$-value $<$ 0.0001). This indicates that channel selection should be paired with feature extraction to lower the computational costs of the decoders, as discussed in previous studies~\cite{PY2019}. The results from the ablation study are presented in Table~\ref{Table 3}. Overall, the study effectively demonstrates the significance of each step employed in the \textit{EEGForceMap} for improving model accuracy.

\subsection{Limitations}

The proposed models demonstrate promising results, but two critical shortcomings need to be addressed. First, the EEG data from the WAY-EEG-GAL dataset focuses on motor execution trials~\cite{luciw2014multi}, leading to strong mu band activation (Fig.~\ref{Figure 3}) in the premotor area associated with movement~\cite{ds09}, leading to the difficulty in applying these models in neuro-rehabilitation. Future studies should develop a dataset that records EEG during grasp force visualisation for application in stroke-rehabilitation protocols. Second, the regressor models show greater errors in decoding rest periods (no force applied) compared to force exertion. This issue may arise from similarities between rest-phase readings and initial force buildup in grasp-and-lift trials. A potential solution is integrating a classifier-decoder ensemble, as suggested by Wu et al.~\cite{10.1007/978-981-99-8021-5_7}, with \textit{EEGForceMap} to first distinguish between rest and task states before regression. Further, these models can be generalised to decode other finger movements like finger strikes using the proposed pipeline. Despite these limitations, the high accuracy in decoding grasp force indicates that the \textit{EEGForceMap} has real-world application potential.

\section{Conclusion} \label{Section 6}

This paper introduces the \textit{EEGForceMap} methodology for EEG preprocessing and feature extraction focused on the neural mechanisms of force control to enhance kinetic decoding accuracy. Among the extracted features, ERDS features correlate most strongly with grasp force, making them ideal for this purpose. The NNR, a deep-learning model integrated with \textit{EEGForceMap}, has achieved the highest accuracy, surpassing current SOTA models. All of the \textit{EEGForceMap} integrated regressors also perform well in both subject-independent and subject-specific contexts, demonstrating its generalisability. Ablation studies confirm that each preprocessing step is essential for optimal performance. While \textit{EEGForceMap} is highly accurate and generalisable, further research is needed to evaluate its real-world effectiveness. All source code and dataset details are available at \url{https://github.com/HAIx-Lab/EEGForceMap} to ensure reproducibility of this work. 

\noindent{Acknowledgment:} This study was supported by the internal research initiation grant, IIT Gandhinagar - IP/IITGN/CSE/YM/2324/05. 




\bibliographystyle{unsrt}
\bibliography{bibtex}

\end{document}